\documentclass[twocolumn,showpacs,preprintnumbers,amsmath,amssymb]{revtex4}

\usepackage{graphicx}% Include figure files
\usepackage{dcolumn}% Align table columns on decimal point
\usepackage{bm}% bold math
\usepackage{float}
\begin{document}

\title{First application of the Trojan Horse Method with a Radioactive Ion Beam:
study of the $^{18}$F$(p,\alpha)^{15}$O reaction at astrophysical energies.}

\author{ S. Cherubini$^{1,2}$,M. Gulino$^{1,3}$, C. Spitaleri$^{1,2}$, 
G.G. Rapisarda$^{1,2}$, 
M. La Cognata${^1}$, L. Lamia${^2}$, R.G. Pizzone${^1}$, S. Romano$^{1,2}$, 
S. Kubono$^{4,5}$, H. Yamaguchi$^5$, S. Hayakawa$^{1,5}$, Y. Wakabayashi$^5$, 
N. Iwasa$^6$, S. Kato$^7$, 
T. Komatsubara$^8$, T. Teranishi$^9$, A. Coc$^{10}$, 
N. de S\'er\'eville$^{11}$, F. Hammache$^{11}$, 
G. Kiss$^{12}$, S. Bishop$^{4,13}$, D.N. Binh$^{5,14}$
}
\affiliation{$^1$INFN-LNS, Catania, Italy}
\email{cherubini@lns.infn.it}
\affiliation{$^2$Dipartimento di Fisica ed Astronomia, Universit\`a di Catania, Catania, Italy }
\affiliation{$^3$Universit\`a di Enna KORE, Enna, Italy}
\affiliation{$^4$Riken, Wako, Saitama, Japan}
\affiliation{$^5$Center for Nuclear Study, The University of Tokyo, Japan}
\affiliation{$^6$Department of Physics, Tohoku University, Sendai, Japan }
\affiliation{$^7$Department of Physics, Yamagata University, Yamagata, Japan }
\affiliation{$^8$Rare Isotope Science Project, Institute for Basic Science, Yuseong-daero, 
Yuseong-gu, Daejeon 305-811, Korea}
\affiliation{$^9$Department of Physics, Kyushu University, Fukuoka, Japan }
\affiliation{$^{10}$Centre de Spectrom\'etrie Nucl\'eaire et de Spectrom\'etrie de Masse, 
  IN2P3, F-91405 Orsay, France}
\affiliation{$^{11}$Institut de Physique Nucl\'eaire, IN2P3,  F-91405 Orsay, France}
\affiliation{$^{12}$Institute for Nuclear Research (MTA-ATOMKI), Debrecen, Hungary}
\affiliation{$^{13}$TUM, Garching, Germany}
\affiliation{$^{14}$30 MeV Cyclotron Center, Tran Hung Dao Hospital, Hoan Kiem District,
Hanoi, Vietnam}

\begin{abstract}
Measurement of nuclear cross sections at astrophysical energies involving unstable species 
is one of the
most challenging tasks in experimental nuclear physics. The use of indirect methods is often 
unavoidable in this scenario.
In this paper the Trojan Horse Method is applied for the first time to a radioactive 
ion beam induced reaction studying the $^{18}$F(p,$\alpha$)$^{15}$O process
at low energies relevant to astrophysics via the three body reaction 
$^{2}$H($^{18}$F$,\alpha^{15}$O)n.
The knowledge of the $^{18}$F$(p,\alpha)^{15}$O reaction rate is
crucial to understand the nova explosion phenomena. 
The cross section of this reaction is characterized by
the presence of several resonances in $^{19}$Ne and possibly interference 
effects among them. 
The results reported in Literature are not satisfactory and new investigations 
of the $^{18}$F(p,$\alpha)^{15}$O reaction cross section will be useful.
In the present work the spin-parity assignments of relevant levels have been discussed
and the astrophysical $S$-factor has been extracted considering also interference 
effects. 
\end{abstract}
\maketitle

%\section{Introduction}
In many astrophysical scenarios radioactive nuclei 
are fundamental in two aspects: directly because of their role in the nucleosynthesis 
path and 
indirectly as information carriers that allow to develop and check astrophysical models.\\
Unfortunately, the difficulties typical of measurements of nuclear reaction cross sections
at astrophysical energies greatly increase 
when using Radioactive Ion Beams (RIBs). This is mainly due to the available intensities 
of these beams, at least three orders of magnitude lower than the stable ones.
In this framework, the development and the use of reliable indirect methods to measure 
these cross sections becomes even more important than in the case of experiment with 
stable beams.\\
The Trojan Horse Method (THM) has been widely developed and exploited over the past 
two decades to study bare nuclear cross sections at very low energy and it 
is now routinely used to study reactions between charged particles even at zero energy. 
A detailed description of the 
method can be found elsewhere \cite{Spitaleri_Folgaria90, Cherubini96, Lamia_Spitaleri04,
Baur_Typel04,Tumino07,Tumino08,Akram08,Gulino10,Lacognata11,Lamia12,Gulino13nn,Gulino13}
and goes beyond the scope of this paper: a review of the method can be found 
in \cite{Spitaleri_RPP}.
Here we present a pioneering work where the THM has been applied for the first time
to a reaction induced by a RIB, namely the $^{18}$F(p,$\alpha$)$^{15}$O process at nova energies.\\
The $\gamma$-ray emission following the nova explosion is dominated by the $511\,$ keV energy line, 
coming from the annihilation of positrons produced by the decay of radioactive nuclei. 
Among them, $^{18}$F is especially important because of its expected abundance 
in the Nova environment and because of its lifetime, that matches well with the timescale 
for the Nova ejecta to become transparent to $\gamma$-ray emission. In order  
$\gamma$-ray astronomy to be helpful in understanding the Nova explosion phenomena, it is then crucial 
to know the rate of the nuclear reactions producing and destroying $^{18}$F. Indeed, at relevant 
temperatures (T$_9 \approx 0.2-0.4$), which corresponds to Gamow windows energies in the 
center-of-mass of $E_\textrm{c.m.} \approx 100-400$ keV, 
the $^{18}$F(p,$\alpha$)$^{15}$O 
reaction is expected to dominate over the $^{18}$F(p,$\gamma$) by roughly a factor 
of 1000 and it is the uncertainty in this reaction rate that gives the main nuclear 
contribution to the overall uncertainty in the final abundance 
of $^{18}$F \cite{coc2000,Diehl2013}.  
\begin{figure}[t]
\begin{center}
{\includegraphics[scale=0.33,
]{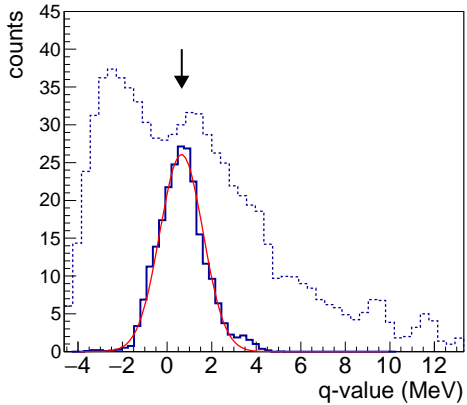}}
\end{center}
\caption{\label{qvalue} Q-value spectra. The Q-value spectrum for the 
$^{2}$H($^{18}$F$,\alpha^{15}$O)n 
reaction (solid line) is overlaid to the total Q-value spectrum (dashed line). 
The arrow represents the
theoretical Q-value (0.658 MeV) for the reaction at hand. The line is a Gaussian fit 
with $\mu$=0.668 MeV and 
$\sigma$=0.322 MeV, in agreement with the expected theoretical value and the 
experimental resolution.}
\end{figure}
In the energy range of interest the cross section of this reaction is dominated 
by the contribution from states in the $^{19}$Ne compound nucleus
around the $^{18}$F+p threshold. 
Especially important is the knowledge 
of the width and spin-parity of these resonance states, in order to verify 
possible interference effects affecting the rate. 
However, in spite of many and long lasting attempts to infer information on the  
$^{18}$F(p,$\alpha$)$^{15}$O reaction by both direct and indirect measurements (see e.g. 
\cite{wiescher1982,rehm1995, coszach1995, rehm1996, rehm1997,graulich1997, utku1998,
butt1998,graulich2001,bardayan2001,bardayan2002,kozub2005,desereville2003, 
nicola2009,beer2011,adekola2011,alison2013}), the situation is still not satisfactory. 
In order to get new complementary information on this process, 
the $^{18}$F(p,$\alpha$)$^{15}$O 
reaction has been studied in inverse kinematics by applying the THM to the 
three body reaction $^{2}$H($^{18}$F$,\alpha^{15}$O)n.
If one succeeds in selecting the phase space region where this reaction proceeds 
through a quasifree reaction mechanism, then its cross section can be factorized 
\cite{Spitaleri_Folgaria90, Cherubini96, Lamia_Spitaleri04,
Baur_Typel04,Tumino07,Tumino08,Akram08,Gulino10,Lacognata11,Lamia12,Gulino13nn,Gulino13}. 
The bare nucleus cross section of 
the $^{18}$F(p,$\alpha$)$^{15}$O reaction can hence be easily deduced even at very 
low energies overcoming the suppression effects coming from coulombian and/or 
centrifugal barriers penetrability factors and the uncertainties due to the poor 
knowledge of spectroscopic information.\\ 
The fundamental step in applying this method is 
the selection of the events that proceed through quasifree reaction mechanism 
among those coming from all other possible mechanisms. To this aim, 
two out of the three reaction products must be detected 
with typical energy and angular resolution of 1\% and 0.1$^{\circ}$-0.2$^{\circ}$ 
respectively. In particular, 
some reconstructed kinematical variables (like relative energies of the outgoing particles) 
are expecially 
sensitive to the angular resolution of the experimental setup. 
Given the characteristics of the radioactive beam used in the present experiment 
and to achieve the angular resolution required by the THM, 
the experimental setup \cite{Gulino13nn} was designed so that it was possible to 
track the beam particles
event by event and to detect ejectiles going in a large solid angle.
\begin{figure}[t]
\begin{center}
{\includegraphics[scale=0.38]{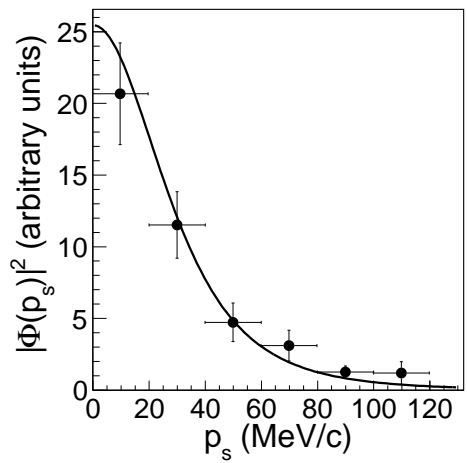}}
\end{center}
\caption{\label{ps} Momentum distribution for the $p$-$n$ intercluster
motion in deuteron. The solid line is the Hulth\'en function in momentum space.}
\end{figure}
The beam tracking system, based on a pair of Parallel Plate Avalanche Counters (PPACs), allowed 
for an angular resolution of 0.14$^{\circ}$. The two ejectiles were detected by using two 
planes of position 
sensitive detectors covering the angular ranges 2$^{\circ}$-11$^{\circ}$ and 
11$^{\circ}$-31$^{\circ}$ with respect to the geometrical center of the target.
In particular a pair of double-sided multistrip silicon detectors (DSSSD), was used 
to detect the heaviest ejecta. The lighter ones were detected by using the
ASTRHO (A Silicon Array for TRojan HOrse) modular 
system of INFN Laboratori Nazionali del Sud
equipped with 8 bidimensional position sensitive detectors   
(BPSD, 45$\times$45 mm$^2$, 500 $\mu$m thick, spatial resolution 1 mm) similar 
to those described in  \cite{Motobayashi_bpsd_89,Motobayashi_bpsd_90}. 
For each detector the $x$ and $y$ position and the energy of the impinging particles were 
recorded together with the time of flight. A standard energy calibration procedure was 
used for all detectors while the x-y position coordinates given by the BPSD were 
calibrated by placing grids in front of them.\\
The emission angles of the ejectile tracks with respect to the measured beam track 
were reconstructed with an overall angular resolution of 0.2$^{\circ}$.\\
The $^{18}$F beam was produced via the $^{18}$O(p,n)$^{18}$F reaction using the CNS
Radioactive Ion Beam (CRIB) separator of the Center for Nuclear Study (CNS) of the University of 
Tokyo, installed at RIKEN accelerator facility in Wako, Japan \cite{yanagisawa2005, yamaguchi2008}. 
The characteristics of the $^{18}$F beam during the experiment were the following: 
beam energy peaked at $47.9\, MeV$  (FWHM $1.9 \,MeV$), maximum beam intensity of
2$\times$10$^6$ pps and purity better than 98\%. The intensity was higher 
than 5$\times$10$^5$ pps throughout the measurement.
The secondary target was made of a thin (typically 150 $\mu$g/cm$^2$) CD$_2$ foil.
The chosen beam energy fulfills the prescriptions for
the validity of the Impulse Approximation \cite{wick52}: (a) the wavelength associated to the entry 
channel is smaller than the nuclear radius of deuteron ($\lambda=1.54$ fm), 
in order to maximize the probability 
that quasifree reaction mechanism occurs; 
(b) the incident center-of-mass energy (4.82 MeV) is higher than the binding energy 
of deuteron.\\
The used setup assures that the coincidence efficiency 
is independent from the excitation energy. A detailed Monte Carlo simulation was  
performed for the used geometry and kinematics 
assuming that the $\alpha$+$^{15}$O breakup is isotropic in the center-of-mass system. 
The actual beam energy profile, the reconstructed beam tracks distribution and 
the measured beam position on the target 
were used in the simulation. The result of the simulation was used to calibrate 
the solid angle coverage of the detectors. 
In order to select events coming from the 
reaction of interest, several cuts were applied on 
the data set. 
In particular it was requested that only one of BPSD and one strip of one 
DSSSDs fired with multiplicity equal to 1 on each detector and that the 
time of flight of the two particle 
was within the correlation time window. 
The events coming from reactions of $^{18}$F on carbon
were easily removed thanks to the difference in phase space distribution of these events 
with respect to those induced on deuterium. 
Moreover, studying the bidimensional spectra of the relative energy of one pair out of three 
outgoing particles versus that of other pairs (e.g. E$_{\alpha-^{15}O}$ vs E$_{\alpha-n}$), 
it was possible to separte the phase space 
region spanned by the events coming from
the $^{2}$H($^{18}$F$,\alpha^{15}$O)n reaction 
that partially overlaps with those coming from other possible reactions on deuterium, 
namely 
$^{2}$H($^{18}$F$,\alpha^{15}$N)p, 
$^{2}$H($^{18}$F,p $^{18}$O)p and $^{2}$H($^{18}$F,p $^{18}$F)n. 
Fig. \ref{qvalue} shows the Q-value spectrum that
was obtained by applying this event selection procedure (solid histogram)
overlaid to the total Q-value spectrum (dashed histogram).
The line shown in this figure is a Gaussian fit 
with $\mu$=0.668 MeV, to be compared with the theoretical Q-value of 0.658 MeV, and 
$\sigma$=0.322 MeV, in agreement with the overall experimental resolution.
The good agreement between the experimental and the theoretical value gives 
confidence in the correct identification of the events coming from the reaction of 
interest.\\
In order to apply the THM, it is essential to select the events coming from the 
quasifree reaction mechanism in the $^{2}$H($^{18}$F$,\alpha^{15}$O)n reaction
among other reaction mechanism.
The strongest evidence of the predominance of the quasifree mechanism 
is given by the shape of the momentum distribution for the $p$-$n$ intercluster
motion in deuteron \cite{Gulino10}. Data are shown in Fig. \ref{ps} by black
dots. The solid line in Fig. \ref{ps} represents the
Hulth\'en function in momentum space with standard parameters values \cite{zadro89}.
The fair agreement between data and theoretical curve give confidence 
in the selection of events coming from the quasifree reaction channel and allows to use
a Plane Wave Impulse Approximation in the data analysis \cite{zadro89}. Indeed, 
it is well known that distortion effects 
influence the behaviour of the momentum distribution at higher values of momentum.
Consequently, the rest of the analysis was done choosing the events having 
a spectator momentum lower than 60 MeV/c. 
\begin{figure}[t]
%\centering
% \resizebox{12.0cm}{!}
\begin{center}
{\includegraphics[scale=0.4]{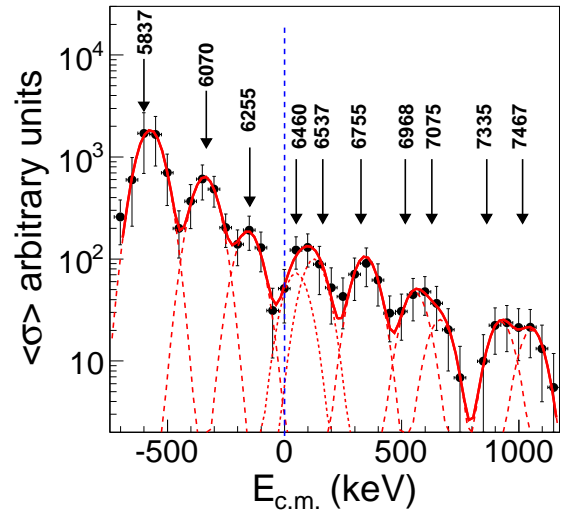}}
\end{center}
\caption{\label{xsec} The nuclear cross section spectrum in function of the 
p-$^{18}$F cm energy for the events that pass the conditions described in the text. The 
blue vertical line shows the position of the threshold for the $^{18}$F+p reaction 
($E_{th}=6.41$ MeV). The red dashed lines represent the gaussians used for fitting the data
as explained in the text. The numbers above the arrows represents the peak positions in 
$^{19}$Ne excitation energy obtained from the fitting procedure.}
\end{figure}
Assuming that the events selected according to the previous procedure come from the 
quasifree contribution to the reaction yield, the 
cross section of the $^{18}$F(p,$\alpha$)$^{15}$O process can be deduced dividing the 
yield of the $^{2}$H($^{18}$F$,\alpha^{15}$O)n reaction by a kinematical factor and 
the momentum distribution of $p$ and $n$ inside the deuteron \cite{Spitaleri_Folgaria90,
Cherubini96, Lamia_Spitaleri04,
Baur_Typel04,Tumino07,Tumino08,Akram08,Gulino10,Lacognata11,Lamia12,Gulino13nn,Gulino13,
Spitaleri_RPP}.  
The nuclear cross section
for $^{18}$F(p,$\alpha$)$^{15}$O reaction, unaffected by suppression effects due to 
Coulomb and
centrifugal \cite{Gulino13} barriers, measured by THM down to zero 
(and even negative) p+$^{18}$F relative energy
is shown in Fig. \ref{xsec}.  
The obtained spectra were analyzed using a least-square fit of multiple Gaussian. The 
sigma of the Gaussians was fixed to 
$53$ keV  based on the energy resolution calculated for the present experiment.
The excitation energies coming from the fit together with the error on the peak 
position are listed in Table I.
For comparison energies and J$^{\pi}$ coming from Ref. \cite{nesaraja07,adekola2011,alison2013,
tilley95} are also reported in Table I.
 \begin{table*}[t]
  \centering
  \begin{tabular}{*{11}{l}}
    \hline \hline \\[-1.5ex]
    \bf{ E$_{c.m.}$ (keV)}  & &    \bf{ E (keV)}  & &      \bf{Ref.\cite{nesaraja07} } & &   \bf{ Ref.\cite{tilley95}}  &&  \bf{Ref.\cite{adekola2011}}   & &    \bf {Ref.\cite{alison2013}}  \\
     (present work)  &&   (present work)  & &   C.D. Nesaraja et al.  & &  D.R. Tilley et al.  & &  A.S. Adekola et al.  & &   A.M. Laid et al. \\
  \\[0.2ex]  \hline\hline \\[-1.5ex]
%   &&   	      &&               &&         &&             &&        \\
  -574$\pm$17 & & 5837  	      &&             &&   5832$\pm$9                      &&                                              &&    \\      \hline  
  \\ [-1.5ex] 
                           & &       	      &&                 &&   6013$\pm$7  (3/2,1/2$^-$)  &&           &&      6014$\pm$2 (3/2$^-$)       \\ \hline 
  \\[-1.5ex] 
  \raisebox{-1.5ex}{-341$\pm$16} & & \raisebox{-1.5ex}{6070 }	      &&             &&                                            &&                 &&       6072$\pm$2 (3/2$^+$,5/2$^-$)     \\
  						 & &       	      &&                     &&                                            &&   6089$\pm$2     &&                                                       \\   \hline
   \\[-1.5ex] 
						& &        	      &&             &&  6092$\pm$8                     &&        &&    6097$\pm$3 (7/2,9/2)$^+$        \\ \hline 
 \\[-1.5ex]
%                         & &       	      &&                 &&                     &&                                                &&          \\
                           & &       	      &&                 &&  6149$\pm$20                   &&                                                && 6132$\pm$3 (3/2$^+$,5/2$^-$)            \\  \hline 
 \\[-1.5ex]
  
 -156$\pm$18  & & 6255      &&       &&            &&       &&               \\ 
                            & & 	      &&       &&   6288$\pm$7         &&   6289$\pm$2 (1/2$^+$,3/2$^+$,1/2$^-$)    &&        6289$\pm$3 (5/2,11/2)$^-$        \\  \hline \\[-1.5ex]
 \\[-1.5ex]
%                           & &       	      &&                 &&                     &&                                                &&           \\  %
%                            
                            &&                   &&      6419$\pm$6    (3/2$^+$)     &&                &&   6419$\pm$6  (3/2$^-$)   &&        6416$\pm$3 (3/2$^-$,5/2$^+$)       \\
                            &&                   &&      (6422)$\pm$30   (11/2$^+$)  &&               &&                                &&             \\
                            &&                   &&      6437$\pm$9  (1/2$^-$)      &&   6437$\pm$9 &&     &&        6440$\pm$3 (11/2$^+$)       \\               \\[-1.5ex]                    
					      &&                     &&   6449$\pm$7 (3/2$^+$)  &&          &&                   &&            \\
 49$\pm$14     && 6460  (3/2$^+$, 5/2$^-$)  &&    &&          &&                   &&        6459$\pm$3 (5/2$^-$)     \\ [0.5ex]  
\hline 
\\[-1.5ex]
                                 & &           	&&   (6504)$\pm$30 (7/2$^+$)  &&                       &&                               &&        \\
 126$\pm$15     & &  6537	   (7/2$^+$,9/2$^+$)        &&      &&                             &&                               &&      \\
                                 & &           	&&    (6542)$\pm$30 (9/2$^{+}$)   &&                       &&                               &&        \\  \hline \\[-1.5ex]
                             & & 	                                                   && 6698$\pm$6 (5/2$^+$)       &&                             &&                               &&  6700$\pm$3     \\ \hline \\[-1.5ex]
 344$\pm$18     & & 6755	(3/2$^-$)        &  &    6741$\pm$6  (3/2$^-$)   &&  6742$\pm$7 (3/2,1/2$^-$) &&   6747$\pm$5 (3/2$^-$)   &&  6742$\pm$2 (3/2$^-$)       \\ \hline \\[-1.5ex]
                             & & 	                        &&   (6841)$\pm$30  (3/2$^-$)  &&                              &    &                               &  &      \\ 
                             & & 	                        &&   6861$\pm$6 (7/2$^+$)  &&  6861$\pm$7                             &    &                               &  & 6862$\pm$2 (7/2$^-$)      \\\hline \\[-1.5ex]
                             & &       	       &   &  (6939)$\pm$30  (1/2$^-$)  &    &                           &  &    & &           \\             
 556$\pm$19     & & 6968 (5/2$^+$)	       &   &     &    &                           &  &    & &           \\
                             & &       	       &   &  (7054) $\pm$30  (5/2$^+$)   &    &                           &  &    & &           \\             \hline \\[-1.5ex]
664$\pm$10      & &  7075	(3/2$^+$)      &   & 7075.7$\pm$16 (3/2$^+$)   &&  7067$\pm$9    && 7089$\pm$5 (3/2$^+$)    & &           \\  \hline  \\[-1.5ex]  
                             & &        &   & $\vdots$  &  &  $\vdots$   & &     & &           \\ \hline \\[-1.5ex]
924$\pm$11     & & 7335	       &   &                               &&  7326$\pm$15   &&              \\ \hline \\[-1.5ex]
\raisebox{-1.5ex}{1056$\pm$13}   	  & &  \raisebox{-1.5ex}{7467}      			 &   & 7420$\pm$14 (7/2$^+$)   &&     &&     7431&&           \\ 
      & & 	       &   & 7500$\pm$9 (5/2$^+$)   &&  7531$\pm$15   &&     &&           \\ \\[-1.5ex] \hline \hline
  \end{tabular}
  \caption{Summary of the $^{19}$Ne resonance parameters in the energy range explored 
   by the experiment compared to results from other works. The energies of 
   Ref. \cite{nesaraja07} reported in parenthesis have been not measured.}
 \label{tablivelli}
\end{table*}

\begin{figure}[t]
\begin{center}
{\includegraphics[scale=0.40]{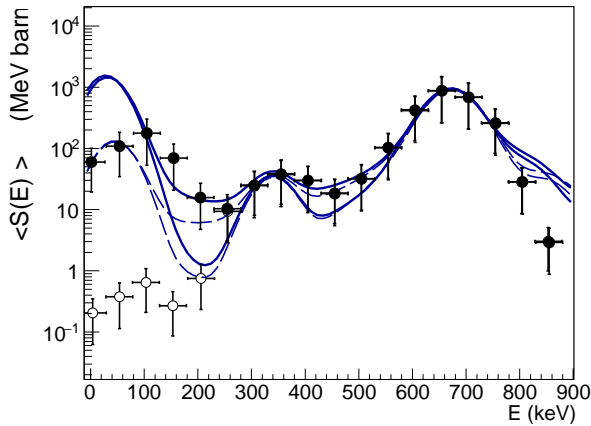}}
\end{center}
\caption{\label{sfac} 
The $^{18}$F(p, $\alpha$)$^{15}$O astrophysical $S$-factor from the present 
experiment. 
The full dots are THM experimental data with the assumption of $J^{\pi}=3/2^+$ for the
resonance at  $E=6460$ keV, the open ones corresponds to the assumption of $J^{\pi}=5/2^-$ 
(the difference from this last assumption to the other possible value 1/2$^-$ 
and 3/2$^-$ being negligeable within the errors). 
The solid and dashed lines shown in figure are calculations presented 
and discussed in Ref. \cite{beer2011} smeared to the present experimental resolution: 
see text for details.
} 
\end{figure}

Further analysis have been done by considering only the energy region of interest 
for astrophysical purposes, i.e. the observed excited states at energies 6255, 6460, 
6537, 6755, 6968 and 7075 keV.\\ 
Though the cross section of the $^{18}$F(p,$\alpha$)$^{15}$O reaction was indirectly 
measured by THM in the center of mass angular range 70${^\circ}<\theta_{cm}<$120$^{\circ}$,
the statistics of the present data do not allow to extract the angular distributions 
of the populated resonances and hence a self-assignment of the J$^\pi$ values is 
not possible in this case.
However, as the THM data are not affected by suppression effects coming from the 
Coulomb and centrifugal barrier \cite{Gulino13, Spitaleri_RPP}, the population of the 
excited states in the compound 
nucleus is linked to the J$^\pi$ of the levels \cite{fuchs1971}.
Hence, for each populated $^{19}$Ne excited states observed in the present experiment
a certain J$^\pi$ value has been assumed by comparison with 
those available in Literature, as discussed below, and it is reported in parenthesis 
in Table I. Using these assumptions, data from each resonance have been integrated over the 
full angular range by means the corresponding Legendre polynomial to obtain the 
contribution to the THM cross section. 
Finally, each contribution to the THM cross section has been multiplied for the 
corresponding penetrability factor
of the centrifugal barrier to get the astrophysical $S$-factor of the reaction.
%%%%%%%%%%%%%%NORMALIZZAZIONE%%%%%%%%%%%%%%%%%%%%%%%%
As THM cannot provide for the absolute normalization of results, the obtained spectrum for 
the $S$-factor was then normalized  to the results of direct measurements of the well 
known resonance at 7075 keV.\\
From Literature, the J$^{\pi}$ values of the two resonances at energies 7075 keV and 6755 keV
are well known to be 3/2$^+$ and 3/2$^-$, respectively, 
and these values have been assumed throughout the present analysis.\\ 
The resonance at $E=6968$ keV has not been observed in previous measurements. Nonetheless,
following Ref. \cite{nesaraja07}, two nearby levels are predicted at 6939 keV 
($J^{\pi}=1/2^-$) and at 7054 keV ($J^{\pi}=5/2^+$). The resonance at $E=6968$ keV could
then be associated with these predicted levels. In this case, fixing a value of
$J^{\pi}=1/2^-$ results in a cross section value that is unlikely bigger than that 
of the resonance at $E=7075$ keV. Moreover,
the fact that the resonance at $E=6968$ keV is not evident in direct measurements  
suggests instead a relatively high $l$ assignment. So $J^{\pi}=5/2^+$ ($l = 2$) 
has been assumed for this level in the present work for the calculation of the 
astrophysical $S$-factor, though we cannot exclude higher
assignments of the value of $l$ for this resonance.\\
With a similar reasoning, the $^{19}$Ne state at $E=6537$ keV 
has been associated with the calculated levels at 6504 keV ($J^{\pi}= 7/2^+$) or 6542 keV 
($J^{\pi}= 9/2^+$) both with $l=2$ \cite{nesaraja07}. 
Both values of the resonance parameters result in a 
strong suppression of the cross section due to the centrifugal barrier penetrability and 
hence do not influence the final value of the astrophysical $S$-factor. Just for computational 
purposes the spin-parity assignment for this level was fixed to be 7/2$^+$.\\
The situation regarding the lower energy region is less clear. 
In the present experiment only a level at 6460 keV is observed.
Above the proton threshold several 
resonances are foreseen by comparing $^{19}$Ne to its mirror 
nucleus $^{19}$F \cite{nesaraja07} and some of them
have been observed \cite{utku1998,adekola2011,alison2013}.  
In particular, a resonance at $E=6459$ keV has been observed in \cite{alison2013} 
and it has been interpreted as part of a triplet of states
with possible spins and parities (3/2$^-$ or 5/2$^+$) at 6416 keV, (11/2$^+$)
at 6440 keV and (5/2$^-$) at 6459 keV.
Calculations presented in \cite{nesaraja07} attribute to four states in the same 
energy region the spin-parity values of 1/2$^-$ (6419 keV), 3/2$^+$ (6422 keV and 
6449 keV) and 11/2$^+$ at the unobserved state at 6422 keV. 
Though only the level at 6449 keV and that at 6459 keV are within the fit error for 
the 6460 keV peak observed in the present work,
calculations of the contribution to the total cross section due to this very level 
have been performed assuming 
all of the $J^{\pi}$ 
values mentioned above. If the spin-parity value is fixed 
to be 11/2$^{+}$ or 5/2$^{+}$ this contribution is strongly suppressed by 
the centrifugal barrier penetrability factor and hence these $J^{\pi}$ assignments
are rejected. On the other hand, there is no reason to rule out the other values 
of $J^{\pi}$ for this level, namely 1/2$^-$, 3/2$^-$, 5/2$^-$ and 3/2$^+$.
Calculations showed that the differences on the contribution of the 6460 keV level to the 
astrophysical $S$-factor for spin-parity assignment 1/2$^-$, 3/2$^-$ 
and 5/2$^-$ are negligeable within the errors. 
To conclude the discussion on the 6460 keV level, it is worth noting that possible 
interference effects in THM are not calculated but are already contained in the 
data \cite{fuchs1971, Spitaleri_RPP}. 
So the other possible assignment 3/2$^+$ to the 6460 keV level will automatically take 
into account interference effects, if any.\\
Finally, in the subthreshold region, the excited state observed here at $E=6255$ keV 
was assigned a $J^{\pi}= 11/2^-$, as already proposed in Ref. \cite{alison2013}: any 
other assignment on the base of the nearby levels observed in \cite{adekola2011, 
alison2013} results in an
unrealistically huge contribution of this level to the total astrophysical $S$-factor.\\ 
In Fig. \ref{sfac} the results obtained in this work for the astrophysical $S$-factor
are presented: assuming $J^{\pi}= 3/2^+$ for the 6460 keV state the result 
is reported as full dots while the $J^{\pi}=5/2^-$ assumption for 
the same level gives the astrophysical $S$-factor shown as open dots. 
In Fig. \ref{sfac} the experimental points are also compared with the 
calculations for the astrophysical $S$-factor presented in Fig. 3 and 4 
of Ref. \cite{beer2011} smeared at the experimental resolution obtained 
in this work. 
In particular, the solid lines represent the upper and lower limits for 
an R-matrix calculation where the interference among the three states 
with $J^{\pi}=3/2^+$  at energies 6419, 6449, 7075 keV has been considered.
The dashed lines represent the same limits for the case with interference between 
the two states at $E=6449$ keV and $E=7075$ keV only. In this latter calculation the authors
of Ref. \cite{beer2011} attributed the value $J^{\pi}=3/2^-$ to the $E=6419$ keV state in 
$^{19}$Ne.\\
Data from the present experiment have been normalized to these
calculations imposing that the integral of the resonance at 
7075 keV is the same in direct and THM data.\\ 
In the energy region below 100 keV, depending on the $J^{\pi}$ assignment chosen for the 
6460 keV level, the present data either agree fairly well with the region given by the 
dashed lines in Fig. \ref{sfac} ($J^{\pi}=3/2^+$, full dots) or  
become much lower ($J^{\pi}=5/2^-$, open dots) than any previous result. 
In both cases the data obtained in this work seems to exclude the existence of three 
interfering states having $J^{\pi}=3/2^+$ represented by the solid 
lines Fig. \ref{sfac}.\\
In conclusion, the THM was applied for the first time to study a reaction induced 
by a radioactive ion beam. Even with the use of indirect methods like THM, the 
measurement of cross sections of interest for nuclear astrophysics remains one of 
the most difficult tasks in nuclear  physics, because the low radioactive ion beam 
intensity adds on the top of the low cross sections typical of astrophysical  
nuclear processes.\\ 
The THM data have been used to obtain the nuclear cross section for 
the $^{18}$F(p,$\alpha$)$^{15}$O reaction and, by comparison with pieces of 
information present 
in Literature, to infer information about the $J^{\pi}$ of 
the $^{19}$Ne nucleus excited states. From this it was possible to
extract the astrophysical $S$-factor for the $^{18}$F(p,$\alpha$)$^{15}$O process. 
In particular resonances in $^{19}$Ne at energies 6255, 6460, 6537, 
6755, 6968, 7075 keV have been observed and studied, as they 
mostly influence the energy region of interest for the nova phenomena.
A value of  $J^{\pi}=5/2^{+}$ has been assigned to the excited state 
at $E=6968$ keV.
For the $E=6537$ keV both $J^{\pi}= 7/2^+$ or  9/2$^+$ are compatible with 
the present data as this contribution to the astrophysical $S$-factor is
negligeable.
In the subthreshold region, the excited state at $E=6255$ keV was assigned 
a $J^{\pi}= 11/2^-$, following \cite{alison2013}.\\ 
Finally, a single resonance has 
been observed just above the proton threshold at $E=6460$ keV.
Different spin assignments have been considered, namely  $J^{\pi}= 3/2^+$ 
(Fig. \ref{sfac} full dots) and $J^{\pi}=1/2^-$, 3/2$^-$ or 5/2$^-$ (Fig. \ref{sfac} 
open dots).
The comparison between the experimental data and the calculations of 
Ref. \cite{beer2011} seems to exclude the presence of two excited states near the proton 
threshold having  $J^{\pi}=3/2^+$. It is worth noting that, if the value 
$J^{\pi}=3/2^+$ is assigned to the $E=6460$ keV state, the effects coming from 
its interference with the $E=7075$ keV state are implicitly present in the THM 
data. Therefore, once measured unambigously the J$^{\pi}$ of this excited state, 
a straightforward indication of the value of the astrophysical $S$-factor -including 
interference effects- in the energy region of interest for Novae could immediately 
come from THM data.\\

%\section{Acknowledgments}
The authors are indebted to the AVF and CRIB technical staff for their 
invaluable contribution to the experiment and to the INFN-LNS workshop 
and electronics laboratory staff headed by Mr B. Trovato and Mr C. 
Cal\`i respectively. Also we thanks Mr. C. Marchetta from the target 
laboratory at INFN-LNS.
This work was supported in part by the Italian Ministry of 
University and Research under Grant No. RBFR082838 (FIRB2008).\\
This work was partly supported by JSPS KAKENHI (Grants No. 21340053 
and No. 25800125).

\end{document}